\documentclass{article}

\usepackage{arxiv}

\usepackage[utf8]{inputenc} 
\usepackage[T1]{fontenc}    
\usepackage{hyperref}       
\usepackage{url}            
\usepackage{booktabs}       
\usepackage{amsfonts}       
\usepackage{nicefrac}       
\usepackage{microtype}      
\usepackage{lipsum}		
\usepackage{graphicx}
\usepackage{natbib}
\usepackage{doi}
\usepackage{amsmath} 
\usepackage{amssymb}  
\usepackage{graphics} 
\usepackage{epsfig} 
\usepackage{eso-pic} 
\usepackage[caption = false]{subfig}

\title{A Spatial Agent-Based Model for Preemptive Evacuation Decisions During Typhoon}

\author{ \href{https://orcid.org/0000-0002-8892-4288}{\includegraphics[scale=0.06]{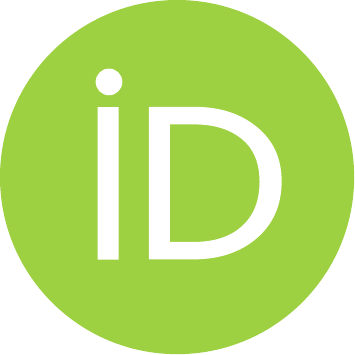}\hspace{1mm}Rey C.~Rodrigueza}\\
	Information and Communications Technology Department\\
	Sorsogon State University - Bulan\\
	Bulan, Sorsogon \\
	\texttt{rey.rodrigueza@sorsu.edu.ph} \\
	\And
	\href{https://orcid.org/0000-0003-0767-9485}{\includegraphics[scale=0.06]{orcid.pdf}\hspace{1mm}Ma.Regina Justina E.~Estuar} \\
	Ateneo Social Computing Science Laboratory\\
	Ateneo de Manila University\\
	Loyola Heights, Quezon City \\
	\texttt{restuar@ateneo.edu} \\
}

\date{}

\hypersetup{
pdftitle={A template for the arxiv style},
pdfsubject={q-bio.NC, q-bio.QM},
pdfauthor={David S.~Hippocampus, Elias D.~Striatum},
pdfkeywords={Agent-Based Model, Typhoon, Preemptive Evacuation Decisions},
}

\begin{document}
\maketitle

\begin{abstract}
Natural disasters continue to cause tremendous damage to human lives and properties. The Philippines, due to its geographic location, is considered a natural disaster-prone country experiencing an average of 20 tropical cyclones annually. Understanding what factors significantly affect decision making during crucial evacuation stages could help in making decisions on how to prepare for disasters, how to act appropriately and strategically respond during and after a calamity. In this work, an agent-based model for preemptive evacuation decisions during typhoon is presented. In the model, civilians are represented by households and their evacuation decisions were based from calculated perceived risk. Also, rescuer and shelter manager agents were included as facilitators during the preemptive evacuation process. National and municipal census data were employed in the model, particularly for the demographics of household agents. Further, geospatial data of a village in a typhoon-susceptible municipality was used to represent the environment. The decision to evacuate or not to evacuate depends on the agent's perceived risk which also depends on three decision factors: characteristics of the decision maker (CDM); capacity related factors (CRF); and hazard related factors (HRF). Finally, the number of households who decided to evacuate or opted to stay as influenced by the model’s decision factors were determined during simulations. Sensitivity analysis using linear regression shows that all parameters used in the model are significant in the evacuation decision of household agents. 
\end{abstract}


\section{Introduction}
Natural disasters are exacting an enormous damage with hundreds of thousands of lives lost in the last decade alone \citep{unisdr2015report}. Asia appears to be the mostly affected continent as in all disaster events, 44\% of it were caused by storms and floods which took 58\% of the total deaths and 70\% of the people affected \citep{annualdisasterstatisticalreview2017}. According to World Risk Index 2018 \citep{worldriskreport2018}, the Philippines is the third country in the world with the highest disaster risk. Almost ninety-nine percent of population in the Philippines is potentially exposed to tropical typhoon, seventy-nine percent of which to the most dangerous class of hazard \citep{pesaresi2017atlas}. This alarming fact necessitates continuously finding solutions to reduce deaths and damage during extreme typhoons and disasters given the fact that the geographical location of the Philippines makes it vulnerable to natural calamities or disasters. In doing so, loss of lives and properties could be lessened, if not totally prevented.  

Understanding how human behave and make decisions during disasters could help in response and rescue efforts. Specifically, understanding factors used to determine whether a unit evacuates or not may help in crafting policy and corresponding implementation plans on how to prepare for disasters and how to properly act and strategically respond during and after a calamity. Some factors that have considerable effect to human behavior during disasters should also be considered. Forewarning, for an instance, can do much to mitigate consequence of disasters.

The Philippines' weather state bureau PAGASA uses its Public Storm Warning System to categorize typhoon intensity and warn the public \citep{pagasa2021}. The Public Storm Warning System has five Public Storm Warning Signals (PSWS) as shown in Table \ref{Table:PSWS}. It also uses a set of color-coded rainfall advisories comprising of 3 colors: yellow, orange, and red. Color yellow advisory warns that flooding is expected in low lying areas and near river channels. Orange advisory tells that flooding is threatening in low-lying areas and near river channels. Finally, red rainfall advisory requires community response as severe flooding is expected. 

\begin{table}[!htbp]
\centering
\caption{Public Storm Warning System}
\label{Table:PSWS}
\begin{tabular}{p{1.1cm}p{1.3cm}p{1.4cm}p{2.8cm}}
\hline
PSWS & Lead Time (hrs) & Winds (kph) & Impacts of the Wind \\\hline
\#1 & 36 & 30 - 60 & No damage to very light damage \\
\#2 & 24	& 61 - 120 & Light to moderate damage	\\                                      
\#3 & 18 & 121 - 170 & Moderate to heavy damage \\
\#4 & 12 & 171 - 220 & Heavy to very heavy damage \\
\#5 & 12 & $>$ 220 & Very heavy to widespread damage \\
\hline
\end{tabular}
\end{table}

Typhoon causes several hazards like strong winds, flooding and landslides due to incessant rain and storm surge prompting at-risk households to evacuate to safe places. Most evacuations during typhoons are conducted in areas that are prone to flooding. Household’s decision to evacuate is based on several significant factors such as characteristics of the decision-maker in the household, capacity-related factors and hazard-related factors \citep{lim2016household, medina2016should, cahyanto2014empirical}. 

Identified decision maker’s attributes includes gender, age, level of education, presence of young children, presence of elderly, presence of pet,  income level, house ownership, length of stay in residence and number of household members \citep{lim2016household, cahyanto2014empirical}. Housing type and previous typhoon experience on one hand are classified as capacity-related factors \citep{whitehead2000heading, hasan2010behavioral, fu2006sequential}. Finally, there are hazard-related factors that affects evacuation decision: distance to typhoon, typhoon forward speed, typhoon wind speed, presence of evacuation message, possibility of flooding, time of the day, source of notice for evacuation and type of evacuation notice received \citep{hasan2010behavioral,fu2006sequential}. 

Results on model estimation for partial and full household-level evacuation decision model \citep{lim2016household} for a large city in the Philippines indicate that there is a higher probability of not evacuating if the head of the family is male as compared to a female which is in consonance with the findings of other studies \citep{horney2010individual,lindell2005household,morrow2005hurricane}; there is a higher probability that the households living in a house with more than one floor level will not evacuate; there is more likelihood that households will evacuate if the level of house damage is high; households who own the house are more likely to evacuate compared to those who are just renting; households with homes made from concrete materials have a higher probability of staying at home compared to households whose houses are made of wood; if the source of evacuation warning came from the authorities then the households are more likely to evacuate compared to when the source of information came from relatives or friends; in the presence of small children, households are likely to evacuate which supports the results of other studies \citep{cahyanto2014empirical, dash2002decision}; and households that are located more than 10 meters away from the source of flood hazard have a high probability of evacuating. 
 
Agent-based modeling (ABM) is an approach to evaluate complex systems where independent and interacting agents makes up its domain. It has also been used to analyze systems that incorporate human behavior during decision-making \citep{macal2005tutorial}. 

Many studies have used agent-based modeling for different purposes. It has been applied to simulate disaster scenarios such as simulation of crowd evacuation in a fire disaster \citep{wagner2014agent}, flood incident management \citep{dawson2011agent}, and, human behavior during earthquake and tsunami \citep{d2014agent, mas2012agent} . However, very little study has been gathered so far that used ABM to understand human behavior during typhoon evacuation. This study attempts to model preemptive evacuation decisions during typhoon. 




\section{Methodology}     
\subsection{Overview of the area under study}
Using ABM, a model was developed for San Vicente, a village in the municipality of Bulan which is located at south-western most tip of the island of Luzon, Philippines. The municipality of Bulan is almost always affected by typhoons, experiencing  an average of three to four typhoons per year. Hazard maps of the municipality shows that it is prone to flooding, and San Vicente is one of the most affected villages during typhoon \citep{hydrometbulan2019}. 

\subsection{ABM Development}
The model environment is a GIS data of San Vicente village. The village is mapped in \textit{OpenStreetMap.org} and then the corresponding shapefiles are downloaded from \textit{extract.bbbike.org}. These shapefiles are collections of road networks, areas, buildings, waterways, and points. The shapefiles are then preprocessed using QGIS 3.4 before importing to GAMA Platform V1.8.0 \citep{taillandier2019building} where the model is implemented. We opted to use GAMA over other modeling platforms to build this model because integrating GIS data in the model using GAMA is seamless. 

\subsubsection{Population Data}
The population data came from two sources: 2015 census and housing data from the Philippine Statistics Authority and local (Municipal) census data and disaster report from Municipal Disaster Risk Reduction and Management Office (MDRRMO) of Bulan.

\subsubsection{Model Description}
\begin{figure}
\centering
\includegraphics[scale=0.5]{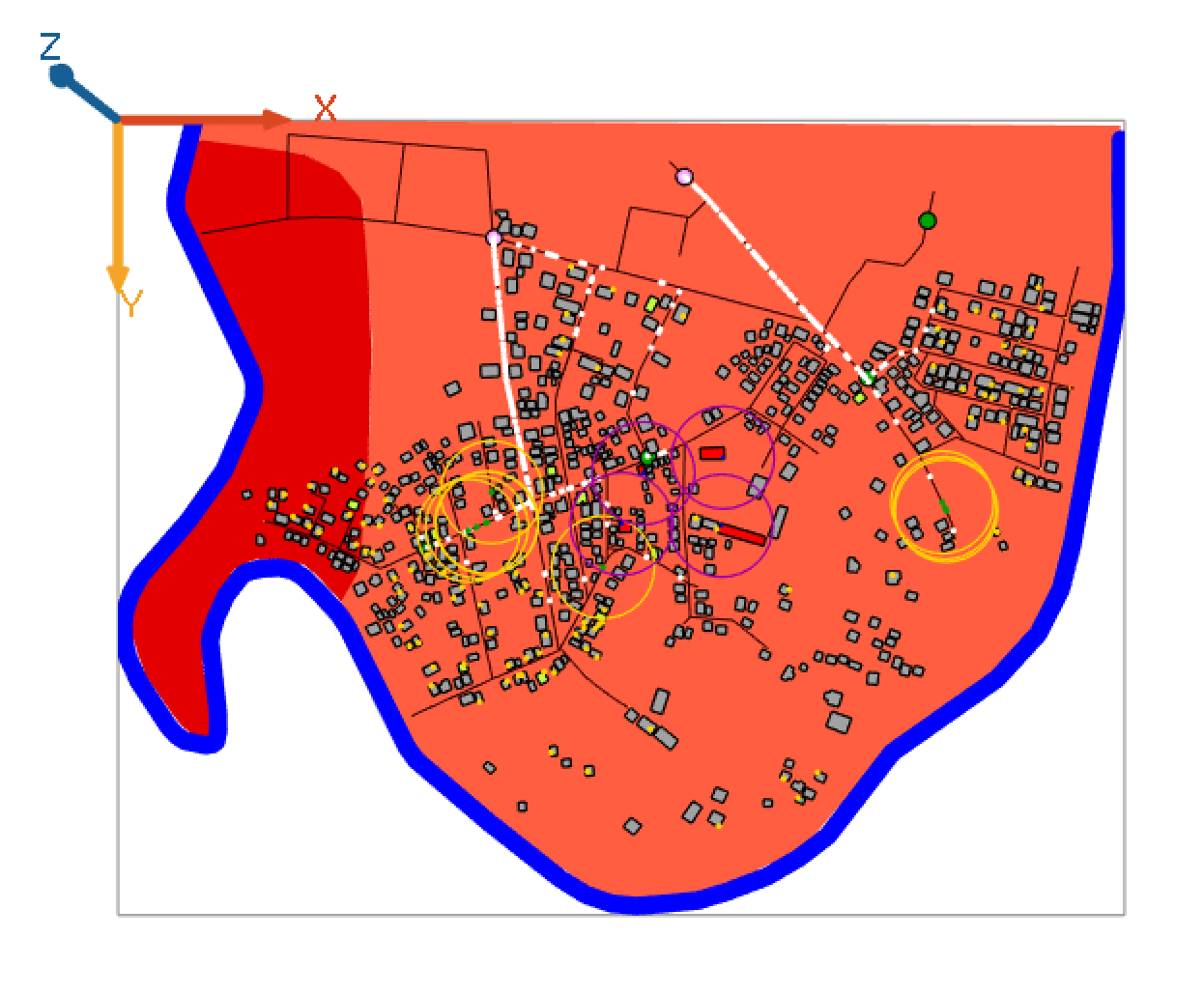}
\setlength{\belowcaptionskip}{-4pt}
\caption{Model simulation showing household agents (white small circles) evacuating.} \label{fig1}
\end{figure}

There are three types of human agents in the model. These are households, rescuers, and shelter managers. The household agents are characterized by the attributes of the head of the household: location, gender, income level, level of education, has small kids in the household, has elderly in the household, has household member with physical disability, type of house ownership, years of residency, past typhoon experience, and perceived risk. Rescuer agents are characterized by their location and capacity. On the other hand, a Shelter Manager agent is characterized by the capacity of the evacuation shelter it is assigned to. Typhoon wind intensity, rainfall severity, and time of day are the exogenous factors/drivers of the model. Each household agent is assigned to a building(house) and each evacuation center has an assigned shelter manager agent. Rescuer agents are traced through their location coordinates. Figure \ref{fig1} shows the model during one of the simulation runs.

\subsubsection{Model Parameters and Initialization}
Table \ref{params} presents the model parameters with their type, definitions and initial values. Other parameters of the model that are used for computing the values of decision factors are presented in Tables \ref{cdmTable}, \ref{hrfTable}, and \ref{crfTable}. The initialization for household agents is always the same and the initial values for rescuers and shelter managers are fixed. The initial values are based on data. These data are sourced from national and local census and interviews from municipal disaster managers.

\begin{table*}[htb]
\begin{center}
\caption{Model parameters and corresponding initial values}
{
\small
\hfill{}
\begin{tabular}{p{2.6cm} p{2.5cm} p{6.5cm} p{1cm}}
\hline
Parameter & Type & Definition & Initial Value \\
\hline
nb\_households & input (constant) & number of household agents & 570\\
nb\_rescuers & input (constant) & number of rescuer agents & 15\\
nb\_sheltermanagers & input (constant) & number of shelter manager agents & 4\\
hhPerDistance & input (constant) & household agent perception distance in meters & 50\\
resPerDistance & input (constant) & rescuer agent perception distance in meters & 50\\
shelPerDistance & input (constant) & shelter manager agent perception distance in meters & 50\\
stormSeverity & input (discrete) & typhoon wind intensity & 1\\
rainfallSeverity & input (discrete) & rainfall intensity & 0.25\\
threshold & input (discrete) & threshold of evacuation decision & 0.7\\
timeOfDay & input (discrete) & time of day (daytime or night time) & 0.5\\
weight\_CDM & input (continuous) & weight for Characteristics of Decision Maker (CDM) decision factor & 0.1\\
weight\_HRF & input (continuous) & weight for Hazard Related Factors (HRF) decision factor & 0.1\\
weight\_CRF & input (continuous) & weight for Capacity Related Factor (CRF) decision factor & 0.1\\
evacuated & outcome & number of household agents who evacuated & 0\\
\hline
\end{tabular}}
\hfill{}
\label{params}
\end{center}
\end{table*}

At time t = 0, 570 household agents are initialized with their characteristics values from a CSV file and then randomly assigned to houses/buildings. There are also 15 rescuer agents randomly placed to their initial positions and 4 shelter managers assigned to four evacuation centers inside the village. Rescuers and shelter manager's parameters are initialized as well. Shapefiles are imported to form the road networks, evacuation points, rescuer's starting points, buildings, as well as water bodies. The order in which household agents decides to evacuate is also random as it depends on their individual perceived risk value. The direction in which the rescuers are headed to inform households about preemptive evacuation is also random.

\begin{table*}[htb]
\begin{center}
\caption{Characteristics of Decision Maker and Coding Format}
{
\scriptsize
\hfill{}
\begin{tabular}{p{2.2cm} p{2cm} p{2.5cm} p{1.5cm} p{2cm} p{2cm}}
\hline
Head of household & \textit{headHousehold} & Level of education & \textit{educLevel} & Income level & \textit{incomeLevel} \\
\hline
Male & 0.5 & College & 0.25 & High & 0.25 \\
Female & 1.0 & High school & 0.5 & Middle & 0.5 \\
       &  & Grade school & 1.0 & Low & 1.0 \\
       &  &  &  &  & \\
\textbf{House ownership} & \textit{houseOwnership} & \textbf{Presence of young children} & \textit{hasChildren} & \textbf{Presence of elderly} & \textit{hasElderly} \\
Owns the house & 0.5 & No & 0.0 & No & 0.0\\
Renting & 1.0 & Yes & 1.0 & Yes & 1.0\\
 &  &  &  &  & \\
 &  &  \textbf{Presence of household member with disability} & \textit{withDisability} & \textbf{Length of stay in the residence} & \textit{yearsOfResidency} \\
 &  & No & 0.0 & $>$ 10 years & 0.5 \\
 &  & Yes & 1.0 & $\leq$ 10 years & 1.0 \\
\hline
\end{tabular}}
\hfill{}
\label{cdmTable}
\end{center}
\end{table*}

\begin{table*}[htb]
\begin{center}
\caption{Hazard-Related Factors and and Coding Format}
{
\scriptsize
\hfill{}
\begin{tabular}{p{2cm} p{1.8cm} p{3.8cm} p{2cm} p{1.5cm} p{1cm}}
\hline
Storm severity & \textit{stormSeverity} & Proximity to the source of hazard & \textit{proximityToHazard} & Time of day & \textit{timeOfDay} \\
\hline
PSWS 1 & 0.25 & Far & 0.25 & Daytime & 0.5 \\
PSWS 2 & 0.5 & Near & 0.5 & Night time & 1.0 \\
PSWS 3 & 1.0 & Within & 1.0 &  & \\
       &  &  &  &  & \\
\textbf{Rainfall severity} & \textit{rainfallSeverity} & \textbf{Source of evacuation warning} & \textit{sourceOfWarning} &  & \\
Yellow & 0.25 & Friends & 0.25 &  & \\
Orange & 0.5 & Media & 0.5 &  & \\
Red & 1.0 & Authorities & 1.0 &  & \\
\hline
\end{tabular}}
\hfill{}
\label{hrfTable}
\end{center}
\end{table*}

\begin{table*}[htb]
\begin{center}
\caption{Capacity-Related Factors and and Coding Format}
{\footnotesize
\hfill{}
\begin{tabular}{p{2cm} p{1.6cm} p{2.5cm} p{1.5cm} p{3.2cm} p{2.5cm}}
\hline
House quality & \textit{houseQuality} & House floor levels & \textit{floorLevels} & Past typhoon experience & \textit{typhoonExperience} \\
\hline
Concrete & 0.25 & $>$1 level & 0.5 & Yes & 0.5 \\
Wood & 0.5 & 1 level & 1.0 & No & 1.0 \\
Light materials & 1.0 &  &  &  &  \\
\hline
\end{tabular}}
\hfill{}
\label{crfTable}
\end{center}
\end{table*}

\subsubsection{Process Overview}
Before the typhoon hits the village, rescuer agents roams around the village and informs the household agents about the impending typhoon, then announces preemptive evacuation.  During this time, household agents computes their perceived risk. If the computed perceived risk is higher than the percentage of evacuation decision threshold to the highest possible perceived risk score then the household agent evacuates to the nearest evacuation shelter inside the village or outside of it. However, if the perceived risk value is lower than the threshold value then the household agent stays (does not evacuate). Shelter Manager agents accepts evacuees in evacuation centers and keeps check on the shelter’s maximum capacity. If maximum capacity of the shelter is reached then incoming evacuees will be directed to other evacuation shelters.

\subsubsection{Individual Decision-Making}
The household agents’ main objective is to survive the onslaught of typhoon. The rescuer agents aim is to achieve zero casualty by informing household agents about the typhoon in advance and announcing preemptive evacuation. Shelter manager agents manages order and capacity of the evacuation center. Household agents computes their perceived risk value in order to make a decision. The perceived risk depends on three decision factors: characteristics of decision maker; capacity-related factor; and hazard-related factor. Bounded rationality is incorporated in the perceived-risk equation by adding a random value from a known range of values. Household agents adapt to exogenous state variables since these variables have direct effect on their risk perception and individual objectives. Typhoon wind intensity, rainfall severity, and time of day are the exogenous factors/drivers of the model. For instance, an increasing public storm warning signal increases the hazard-related factor value which is a component of the perceived risk equation for household agents. In this model, spatial aspects play a role in the decision process of the agents. Since the model environment is a GIS data, the distance between objects is based from the real world. An agent that is near the source of hazard such as river will have a higher risk perception compared to others who are at a distance. Evacuating household agents also will choose the nearest available evacuation center. Additionally, temporal aspects play a role in the decision process as household agents perceives night time as a more risky situation compared to daytime during typhoon. 

\begin{figure}
\centering
\includegraphics[scale=0.4]{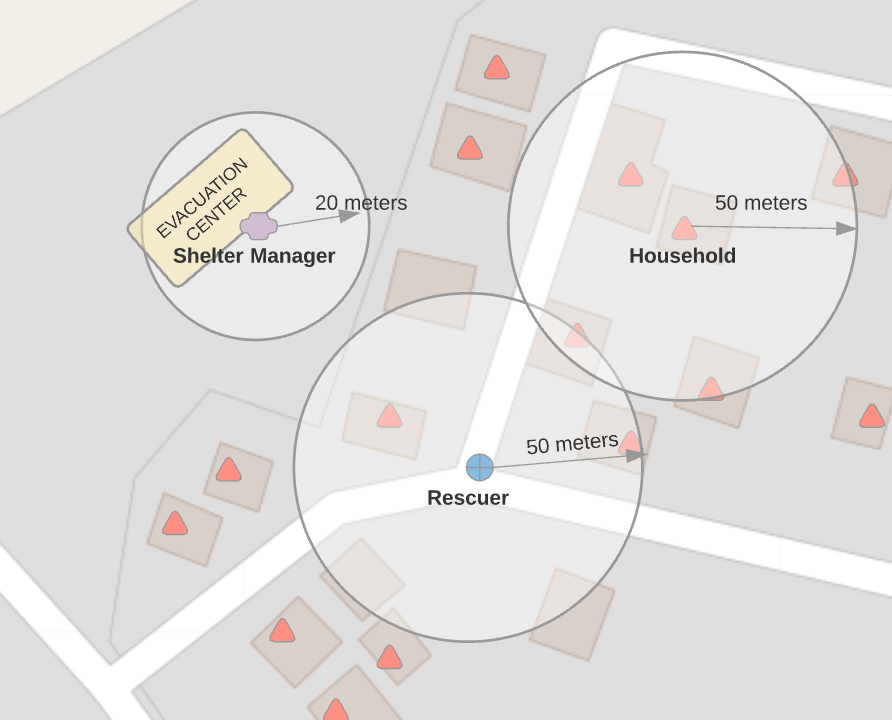}
\setlength{\belowcaptionskip}{-12pt}
\caption{Perception Distances of Household, Rescuer and Shelter Manager Agents.} 
\label{fig:_perception_distances}
\end{figure}

Interactions between agents depend on spatial distances. Each agent type has a perception area in a form of a circle based from perception radius which is set at 50 meters. An agent can be sensed by another agent if it is inside the perception area (which is in a form a circle) of the former. Figure \ref{fig:_perception_distances} shows the perception distances of the three agents in the model. 

\subsection{Computation of the Perceived Risk}
Perceived Risk (PR) of each household agents is based on three decision factors: characteristics of the decision maker (CDM), hazard-related factor (HRF), and capacity-related factor (CRF). PR is computed using equation \ref{eqn:prisk}. Equation \ref{eqn:cdm} computes the value for the characteristics of the decision maker (CDF) in the household. The computation is performed one-time only during simulation. HRF and CRF values are computed by equations \ref{eqn:hrf} and \ref{eqn:crf} respectively.

\small
\begin{multline}
    CDM\ =\ headHousehold\ +\ incomeLevel\ 
    +\ educLevel +\ hasChildren\ +\ hasElderly\ \\ 
    +\ withDisability\ +\ houseOwnership\ 
    +\ yearsOfResidency
\label{eqn:cdm}
\end{multline}
\normalsize
\vspace{-6mm}
\small
\begin{multline}
    HRF = stormSeverity\ +\ rainfallSeverity\ 
    +\ proximityToHazard +\ sourceOfWarning\ \\
    +\ timeOfDay
\label{eqn:hrf}
\end{multline}
\normalsize
\vspace{-6mm}
\small
\begin{multline}
    CRF = houseQuality\ +\ floorLevels\ +\ typhoonExperience
\label{eqn:crf}
\end{multline}
\normalsize
\vspace{-6mm}
\begin{multline}
    Perceived\ Risk = (CDM * weight\_CDM)\ 
    + \ (HRF * weight\_HRF)\ + \ (CRF * weight\_CRF)\ + \  \varepsilon 
\label{eqn:prisk}
\end{multline}
\normalsize

where $\varepsilon$ is the bounded rationality with value between 0.0 to 0.05. The highest possible score for perceived risk is computed as
 \small
\begin{multline}
    Highest\ Possible\ Score = 8\ *\ weight\_CDM\ +\ 3*weight\_CRF\ +\ 5*weight\_HRF
\label{eqn:hps}
\end{multline}
\normalsize

To compute for the highest possible score of perceived risk, the maximum risk value of the factors for each decision factors are added, then multiplied to the assigned weight. For example, CDM decision factors is composed of eight factors where the maximum risk value for each factor is 1.0. These maximum risk values are then added which totalled to 8, then it is multiplied to the assigned CDM weight. The same method is applied for HRF and CDF decision factors. Finally, if household agent's perceived risk value is higher than the percentage of evacuation decision threshold to the highest possible perceived risk score, then that household agent will evacuate. 

\subsection{Experiment Design}
The model have seven experiment input parameters which includes \textit{stormSeverity, rainfallSeverity, weight\_CDM, weight\_HRF,  weight\_CRF}, and \textit{threshold}. Five of the seven parameters have constant increment step values while the other two. \textit{rainfallSeverity} and \textit{timeOfDay} takes on the values {0.25, 0.5 , 1.0} and {0.5, 1.0} respectively. Table \ref{sampledparam} show the minimum and maximum values of the input parameters. There is only one outcome parameter for this model, the parameter \textit{evacuated}. This parameter counts the number of household agents who evacuated upon assessing its perceived risk. 

\begin{table}[htb]
\begin{center}
\caption{Input Parameters}
{\normalsize
\hfill{}
\begin{tabular}{p{2.5cm} p{0.8cm} p{0.8cm} p{0.5cm}}
\hline
Input Parameter & min & max & step \\
\hline
stormSeverity & 1 & 2 & 1\\
rainfallSeverity & 0.25 & 1.0 & \\
timeOfDay & 0.5 & 1.0 & \\
threshold & 0.7 & 0.9 & 0.1\\
weight\_CDM & 0.1 & 0.8 & 0.1\\
weight\_HRF & 0.1 & 0.8 & 0.1\\
weight\_CRF & 0.1 & 0.8 & 0.1\\
\hline
\end{tabular}}
\hfill{}
\label{sampledparam}
\end{center}
\end{table}

The model's parameter space has 18,432 sets of possibilities using exhaustive exploration method: stormSeverity (2) * rainfallSeverity (3) * timeOfDay (2) * weight\_CDM (8) * weight\_HRF (8) * weight\_CRF (8) * threshold (3) = 18,432. Each possibility is replicated 10 times thus generating  184,320 total simulations for the batch experiment. It is to be noted, however, that for a certain parameter combinations or possibility to be considered for simulation run, added weights of CDM, HRF and CRF must total to 1.0.  If the total weight is less than 1.0 then the current possibility simulation is skipped and proceeds to the next one. Of the 18,432 sets of possibilities, only 6,480 are considered and the rest are skipped. 

\subsection{Sensitivity Analysis}
Sensitivity analysis is used to quantify which model inputs are more influential for the model output. To determine significant predictors of evacuation decision, we used the linear regression function \textit{lm()} of the R language on the output data from the batch experiment. The independent variables are the seven sampled input parameters while the dependent variable is the sole model outcome, the number of household agents who evacuated.

\section{Results and Discussion}
Sensitivity analysis shows that all parameters used in the model (weight\_CDM, weight\_HRF, weight\_CRF, stormSeverity, rainfallSeverity, timeOfDay, threshold) are significant in the evacuation decision of household agents having significance levels of $<2e^{-16}$. The $p$-value is also $<2e^{-16}$. Findings also show that the total evacuation decision is more sensitive to threshold and weights assigned in capacity related factors, specifically type of house, floor level and past typhoon experience.

\begin{figure}
\centering
\subfloat[two,orange,nighttime]{\includegraphics[width = 1.7in]{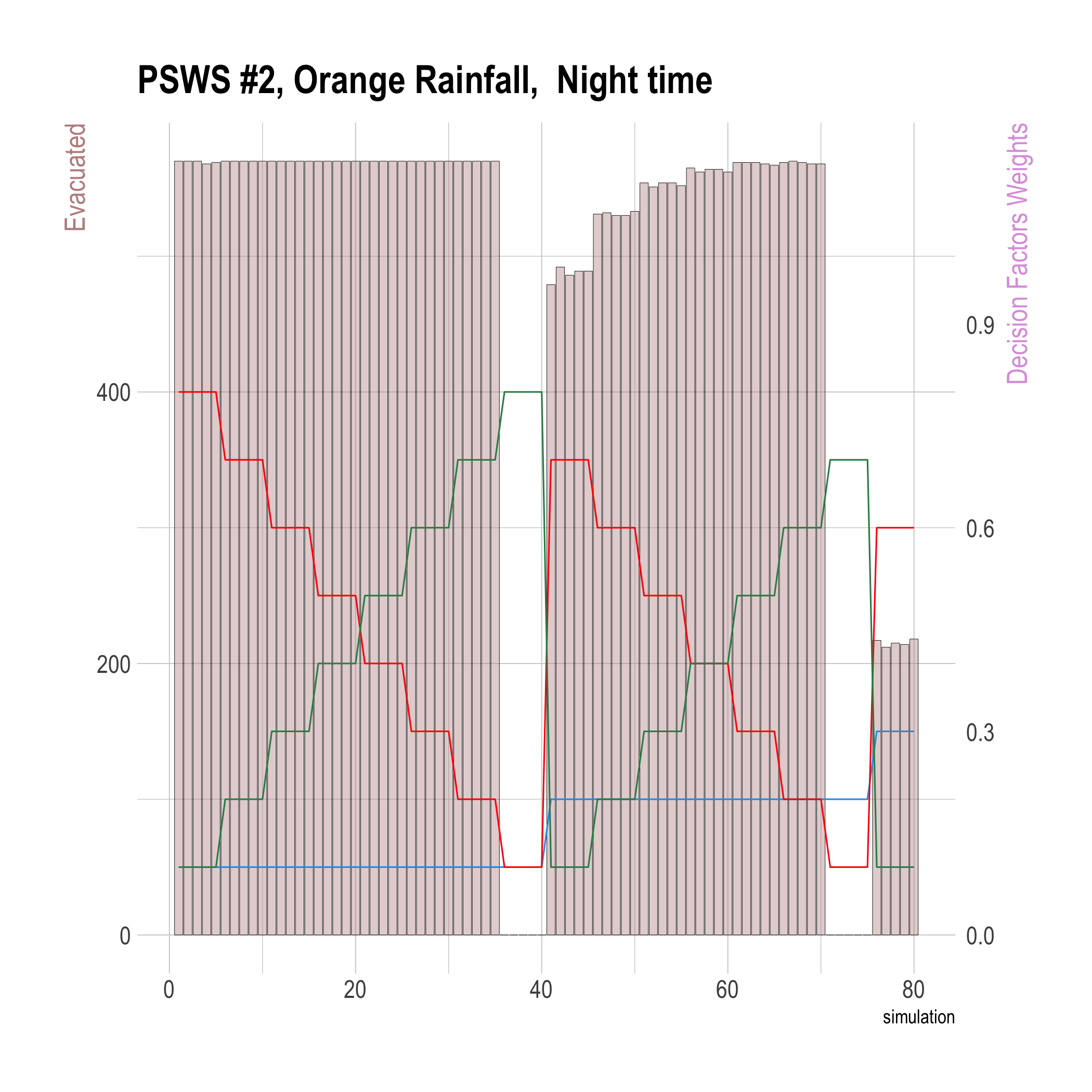}}
\subfloat[two,orange,nighttime]{\includegraphics[width = 1.7in]{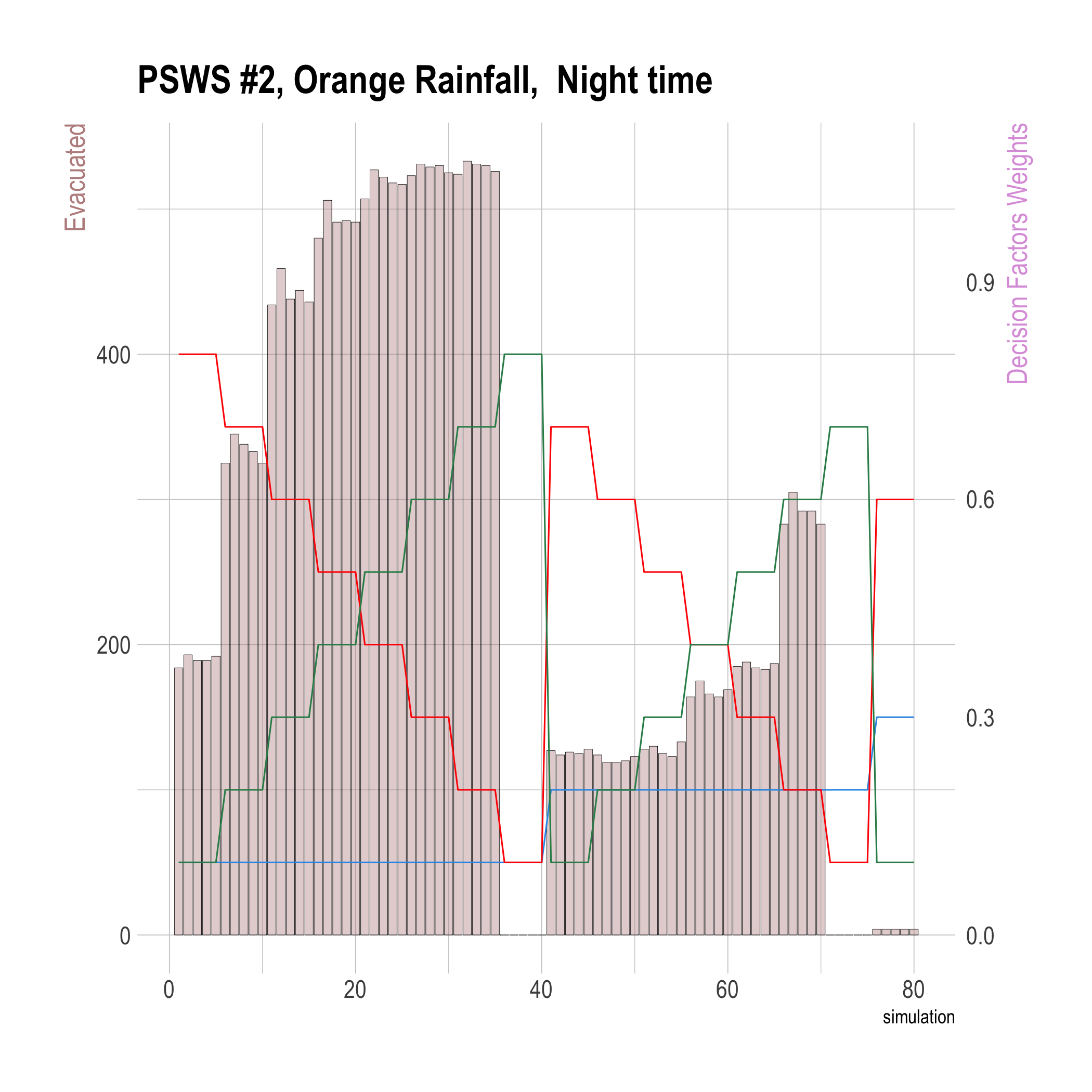}}
\subfloat[two,orange,nighttime]{\includegraphics[width = 1.7in]{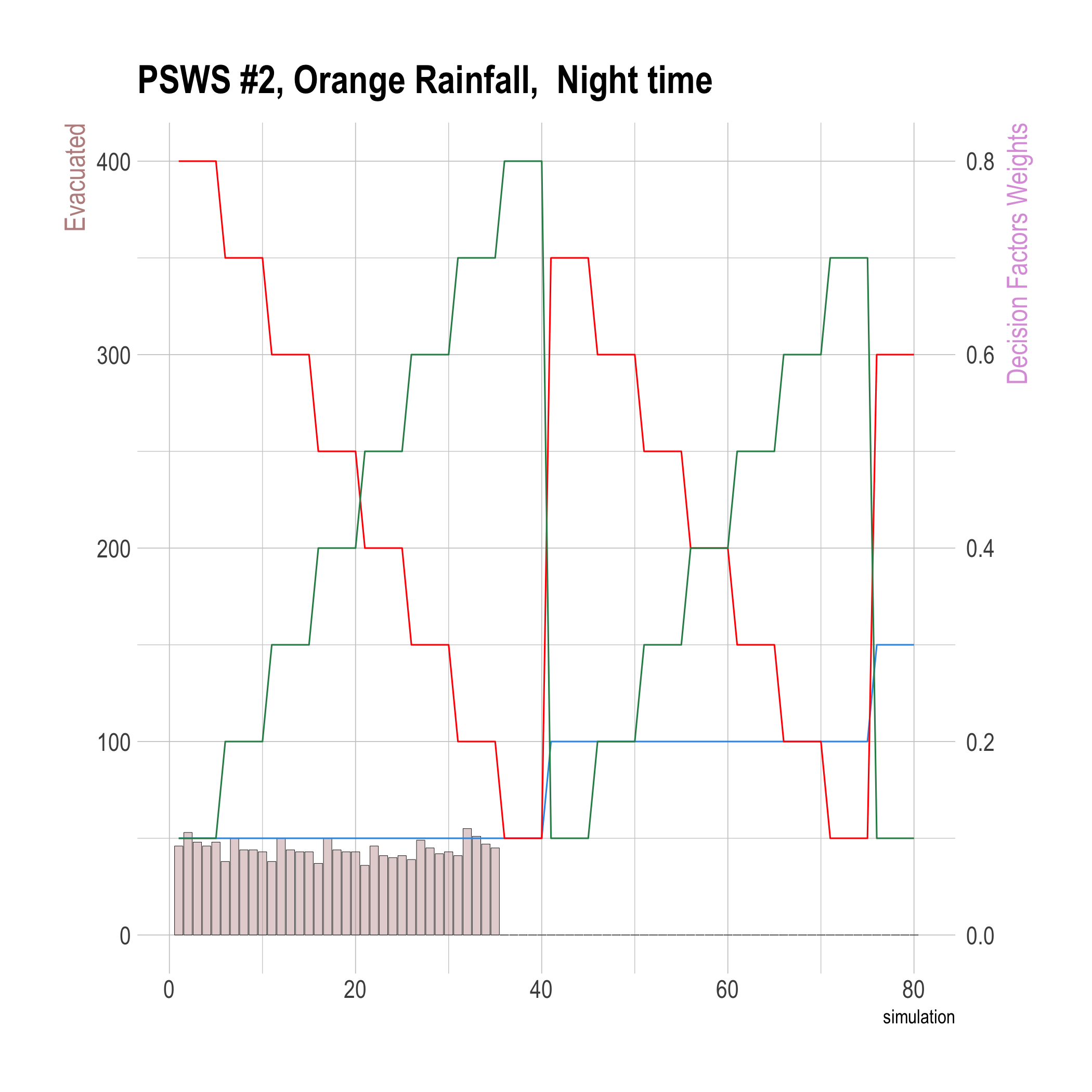}}
\caption{Simulation at (a) 70\%, (b) 80\%, and (c) 90\% threshold of evacuation decision}
\label{fig:result3}
\end{figure} 

Figure \ref{fig:result3} shows selected three simulation results from the batch experiment conducted. The label for each graph follows the format \textit{\{storm severity (public storm warning signal), rainfall severity, time of day\}}. The three colored graph lines represents values for weight\_CRF (red), weight\_HRF (green), and weight\_CDM\_ (blue). The result of one of the three simulation runs, simulation (c), is near to the real-world data based on preemptive evacuation happened during a typhoon that passed in San Vicente village on 2019 where 66 households preemptively evacuated. Simulation (c) result is near to the real-world data based on preemptive evacuation happened during a typhoon that passed in San Vicente village on 2019 where 66 households preemptively evacuated. This indicates that at this typhoon condition, some household decision makers might have reached their evacuation decision threshold and finally decided that the best action is to evacuate. Not doing so would be catastrophic at their end. It is also noteworthy that for all threshold settings, an increase in the weight for capacity-related factors also increases the number of household agents who decided to evacuate. On the other hand, the effect of increasing the weights for characteristics of decision makers and hazard-related factors on the number of evacuees is not evident.

\begin{figure*}
\centering
\subfloat[one,yellow,daytime]{\includegraphics[width = 1.5in]{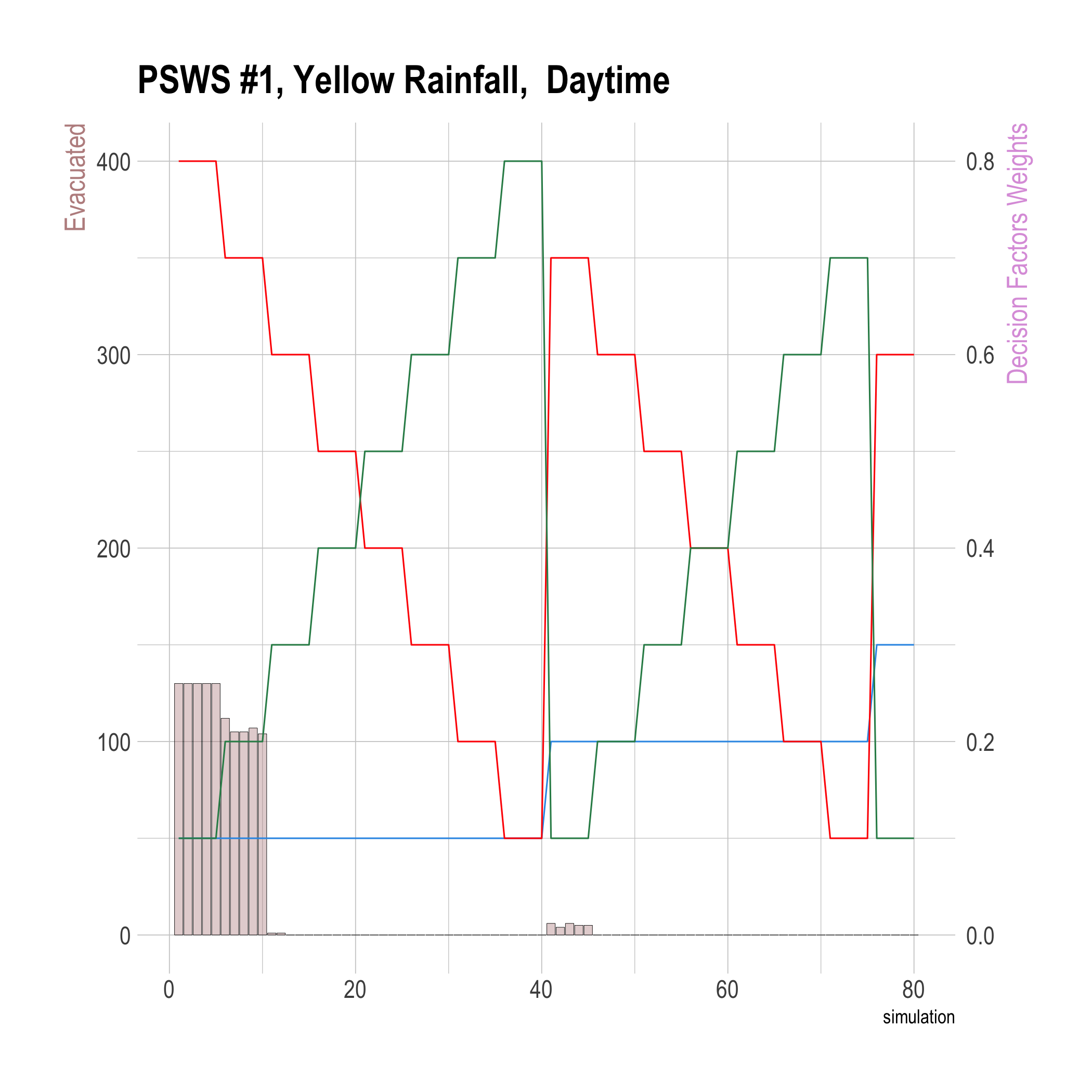}} 
\subfloat[one,orange,daytime]{\includegraphics[width = 1.5in]{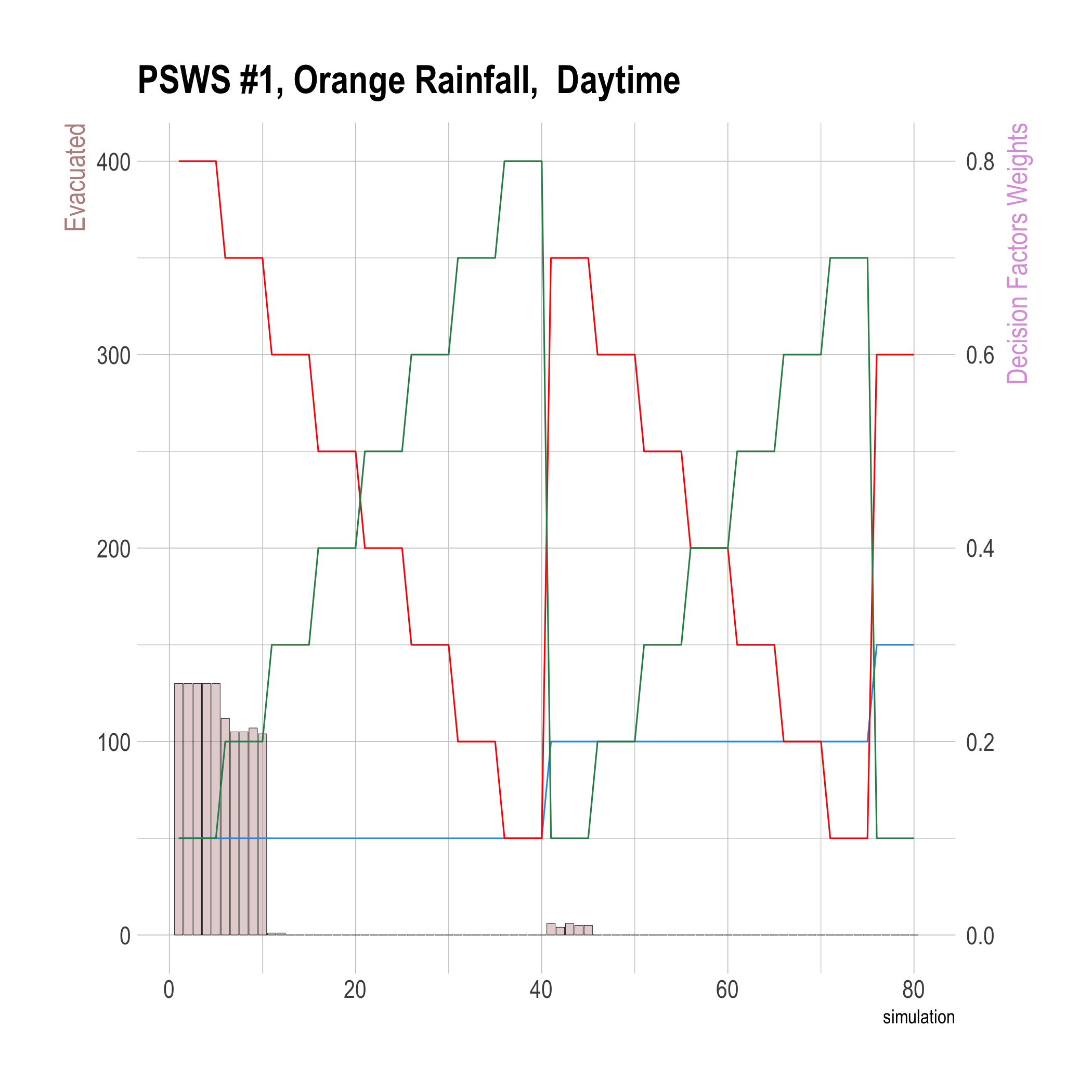}}
\subfloat[one,red,daytime]{\includegraphics[width = 1.5in]{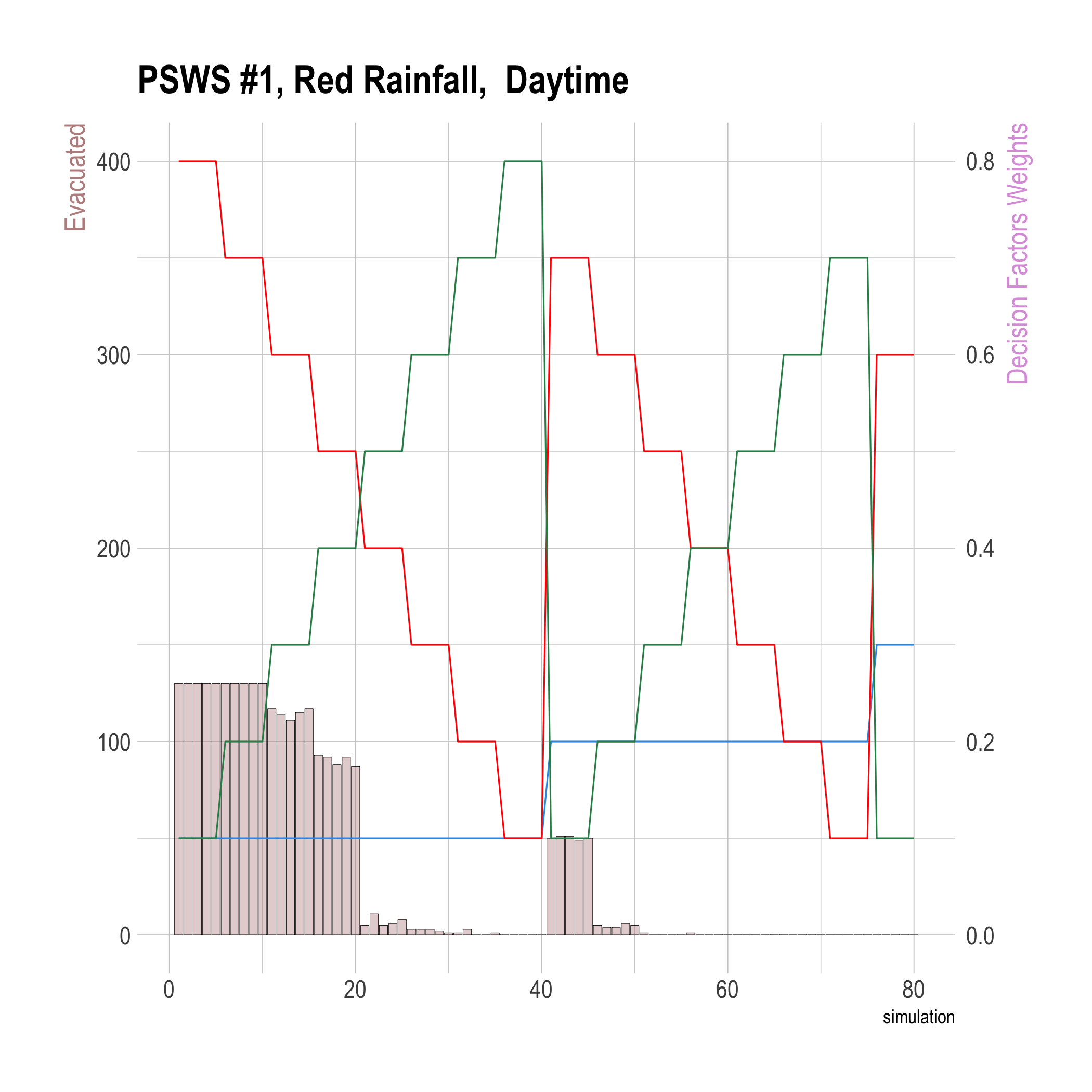}}\\
\subfloat[one,yellow,nighttime]{\includegraphics[width = 1.5in]{graphs80/_1yd.png}}
\subfloat[one,orange,nighttime]{\includegraphics[width = 1.5in]{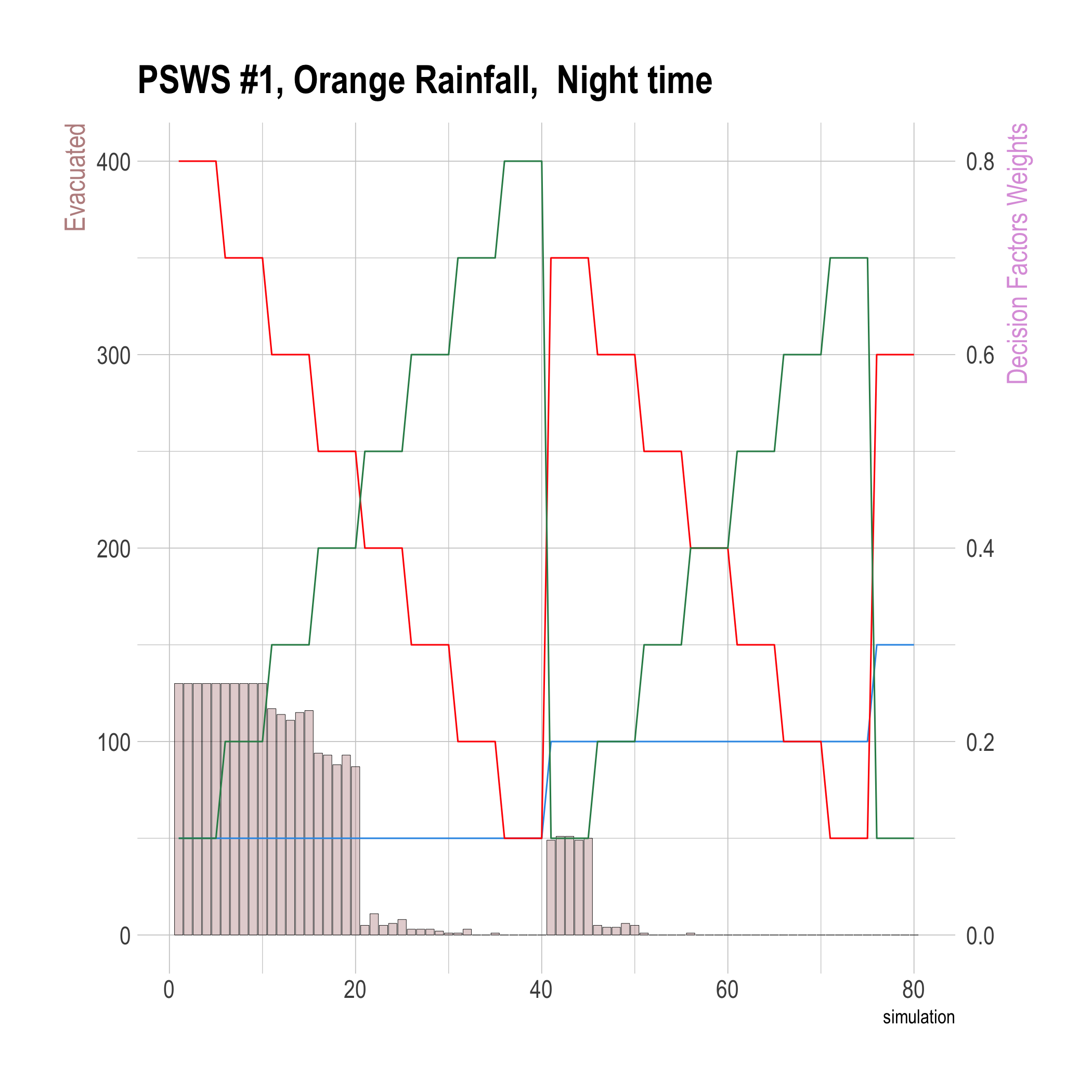}}
\subfloat[one,red,nighttime]{\includegraphics[width = 1.5in]{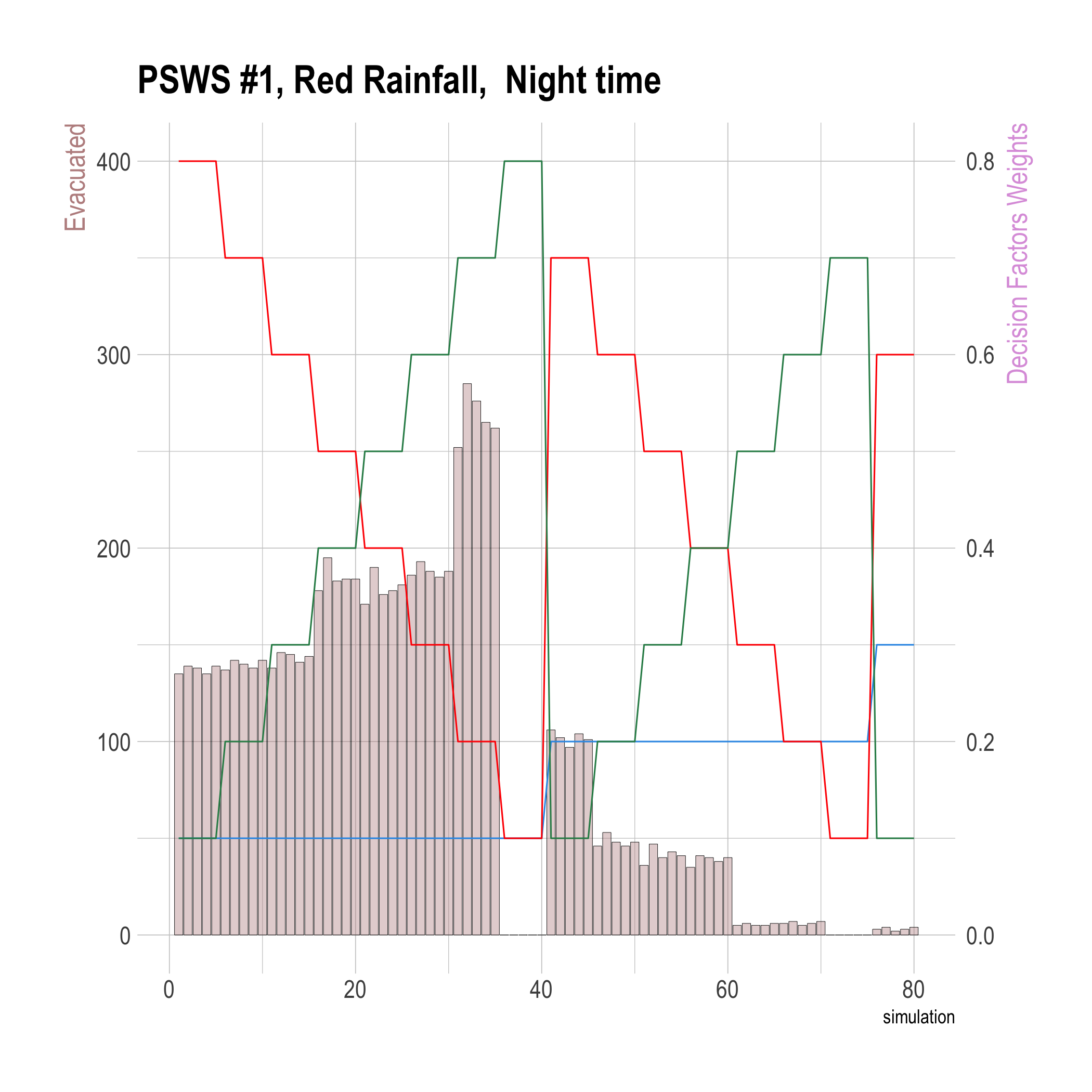}}\\
\subfloat[two,yellow,daytime]{\includegraphics[width = 1.5in]{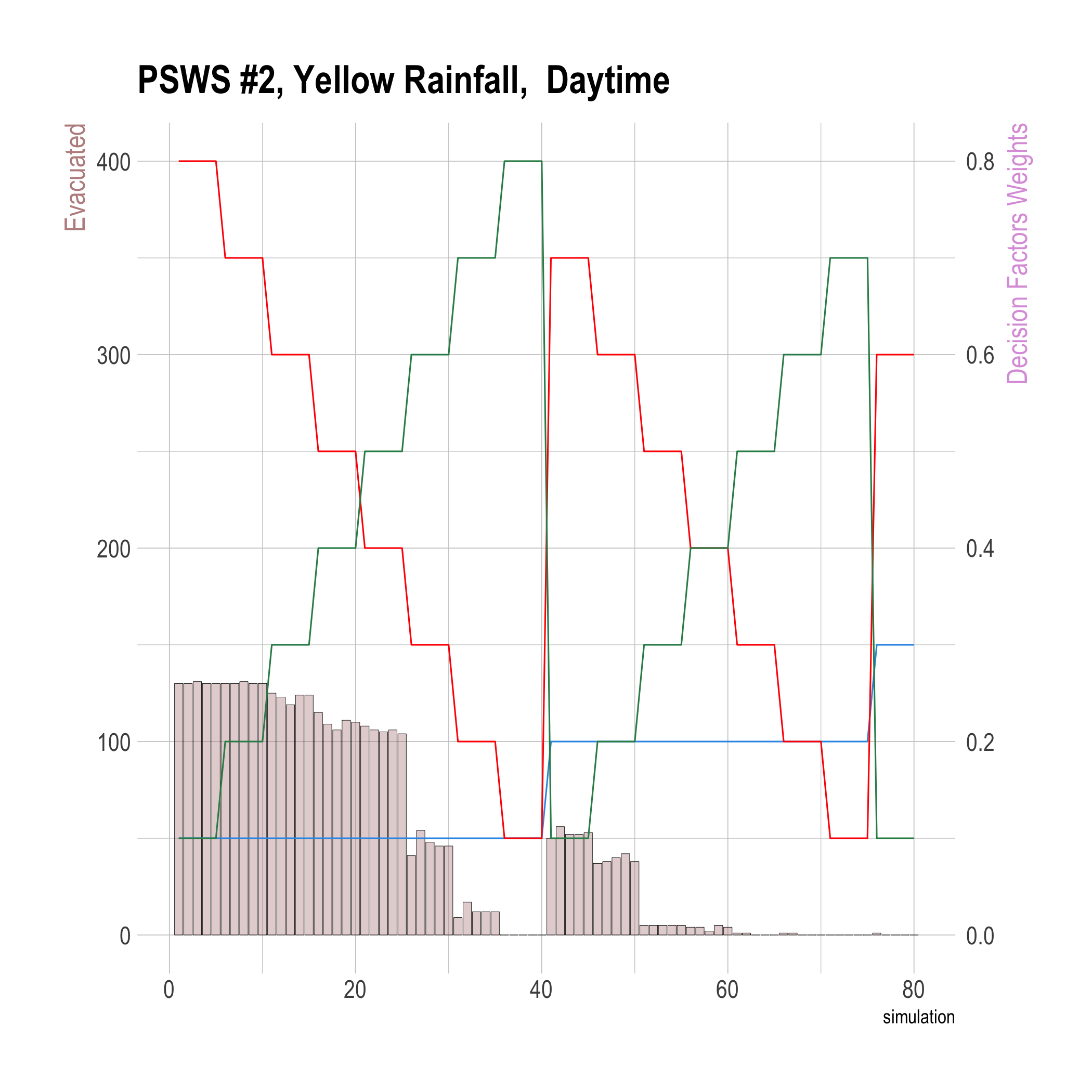}}
\subfloat[two,orange,daytime]{\includegraphics[width = 1.5in]{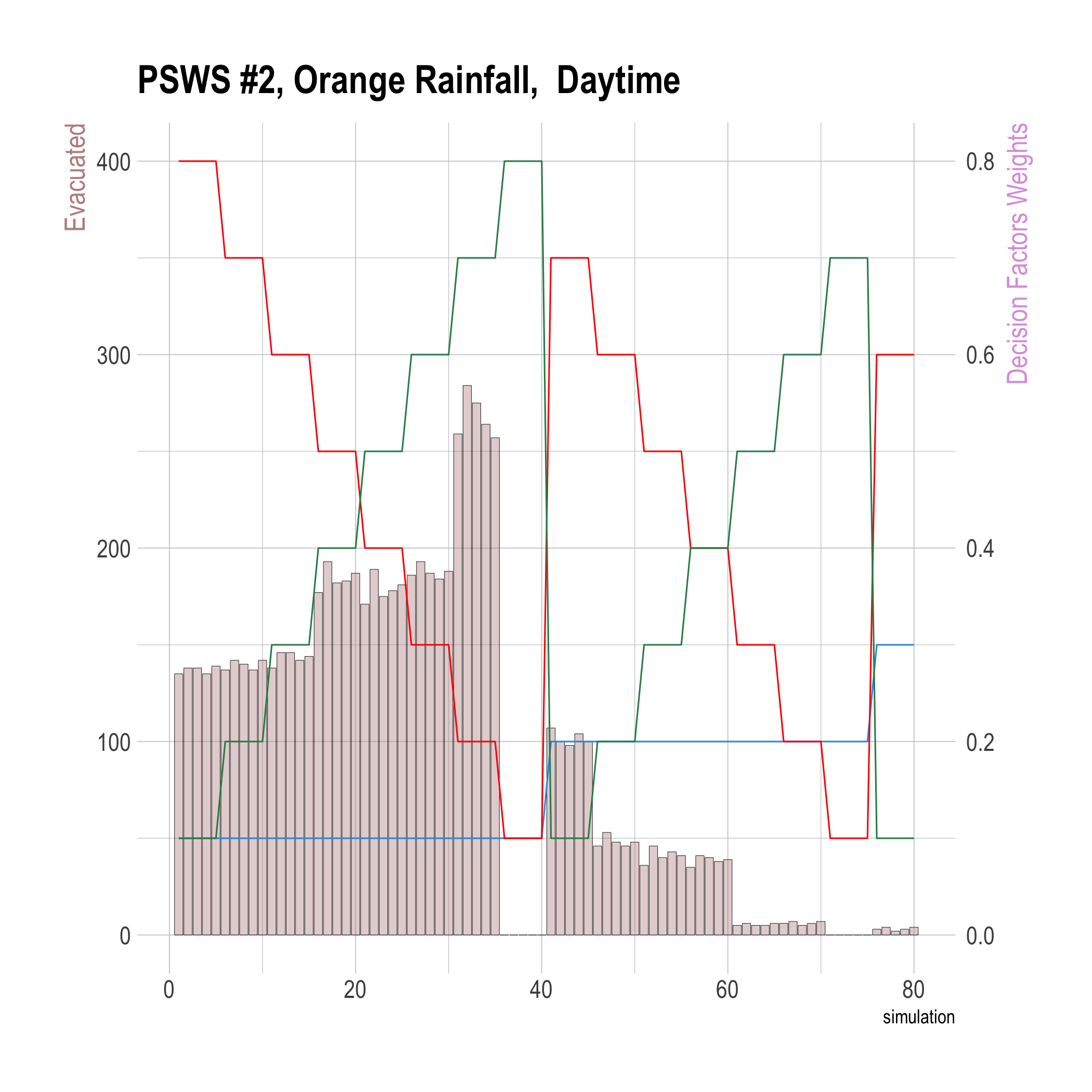}}
\subfloat[two,red,daytime]{\includegraphics[width = 1.5in]{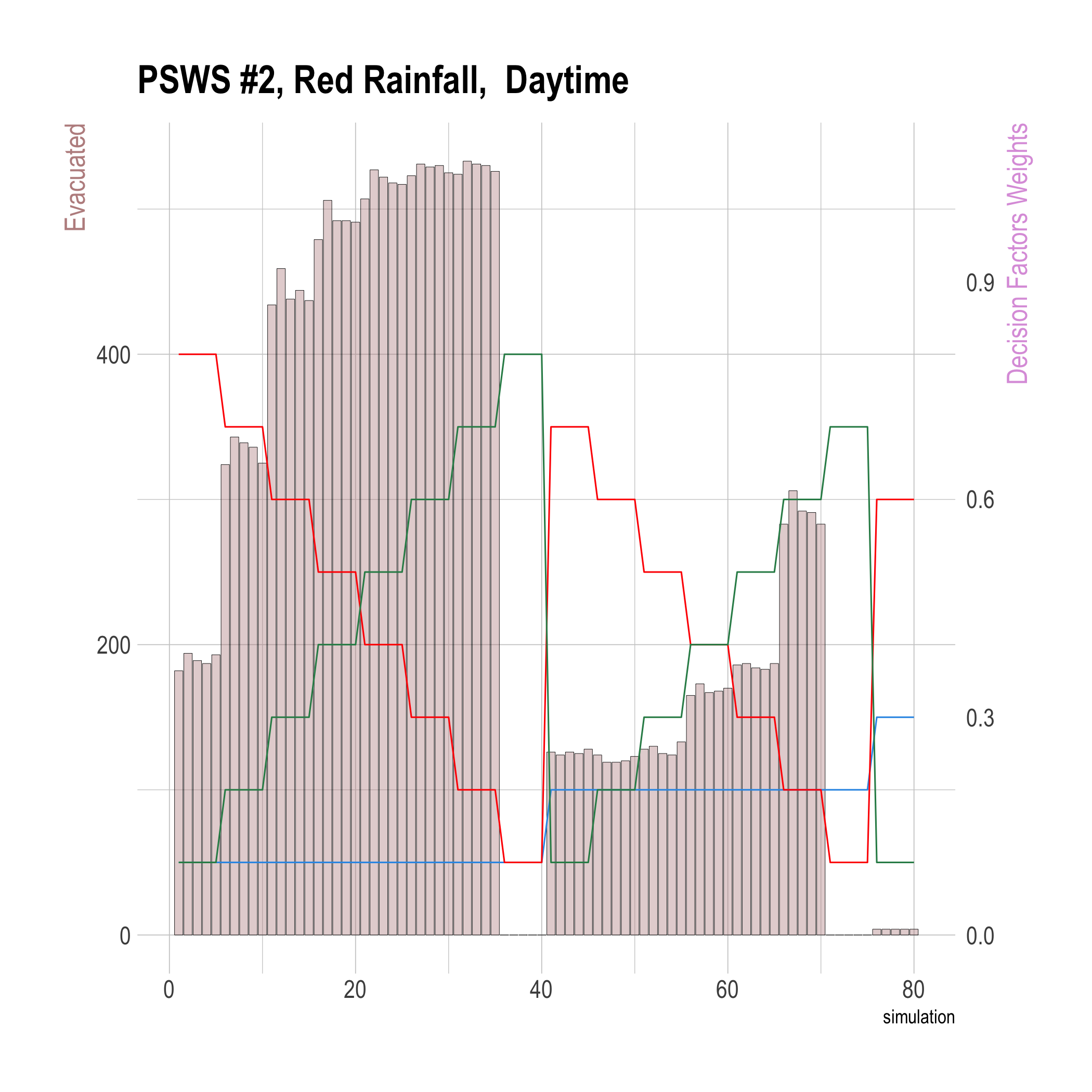}}\\
\subfloat[two,yellow,nighttime]{\includegraphics[width = 1.5in]{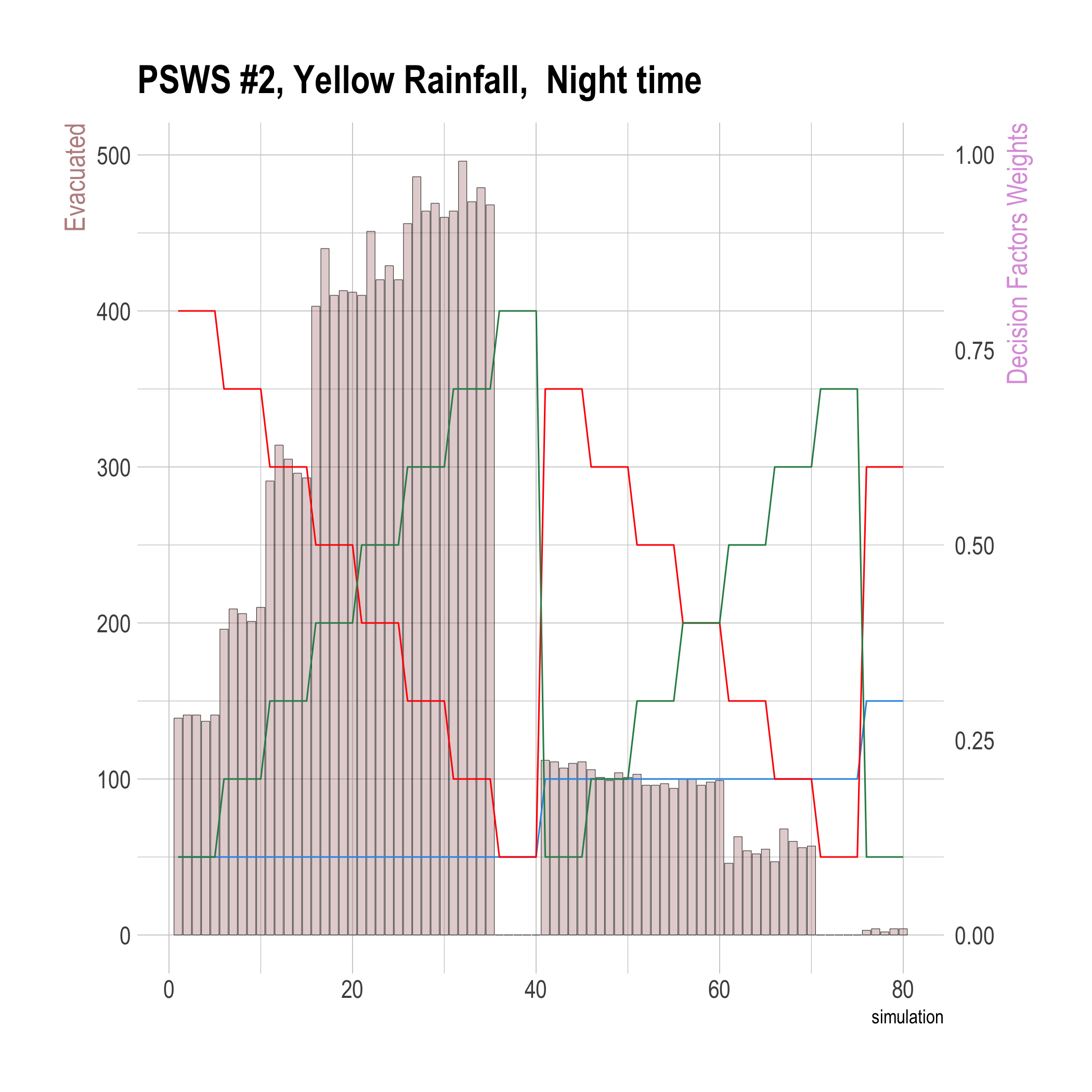}}
\subfloat[two,orange,nighttime]{\includegraphics[width = 1.5in]{graphs80/_2on.png}}
\subfloat[two,red,nighttime]{\includegraphics[width = 1.5in]{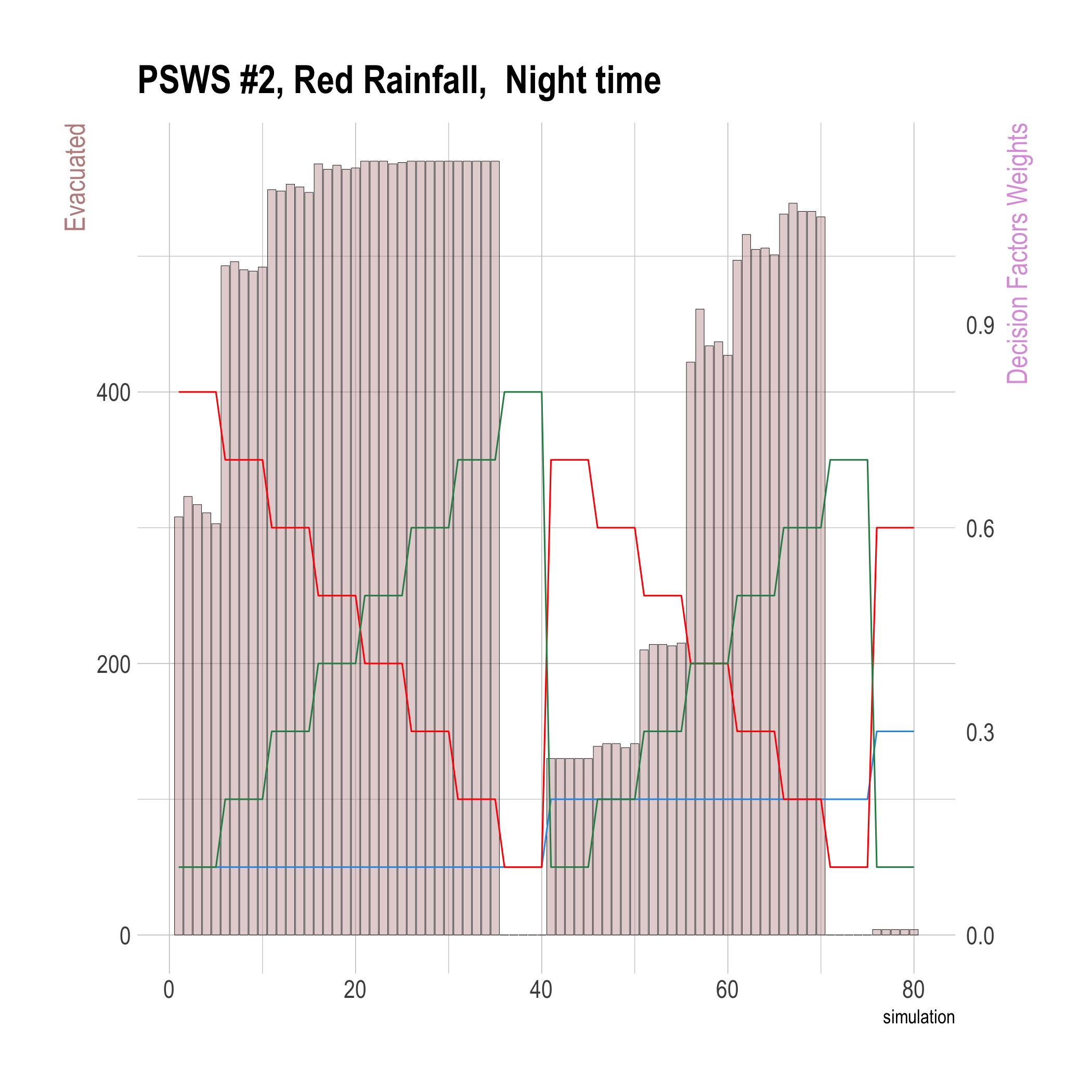}}
\caption{Simulation at 80\% threshold of evacuation decision.}
\label{fig:result80}
\end{figure*}

The batch experiment can actually be segmented into three parts. The first part is where the threshold of evacuation decision is at 70\% and the other two parts are for 80\% and 90\% thresholds respectively. At 70\% threshold, the number of evacuees is high and even reaches to 570 which is the total number of household agents even when the actual risk is low, such as conditions having PSWS 1, yellow rainfall, and day time. The number of evacuees gradually decreased when the threshold is at 80\%, but the numbers are still high compared to real-world data. However, at 90\% threshold, evacuees only started to appear and increase at condition \{two,red,daytime\}. The number of evacuees is at nearest to real-world data at condition \{two,orange,nighttime\}, indicating that at this typhoon condition, some household decision makers might have reached their evacuation decision threshold and finally decided that the best action to do is to evacuate. Not doing so would be catastrophic at their end. It is also important to note that for all threshold settings, an increase in the weight for capacity-related factors also increases the number of household agents who decided to evacuate. On the other hand, the effect of increasing the weights for characteristics of decision makers and hazard-related factors on the number of evacuees is not that evident. Due to limited space, only the simulation result at 80\% threshold is presented in this paper and is shown in Figure \ref{fig:result80}. 

\section{Conclusion}
This paper presented a spatial agent-based model for investigating preemptive evacuation decisions during typhoon. GIS data of a typhoon-susceptible village in south-western part of the Philippines is used as the environment of the model and the data utilized is based on national and local census. Among the three human agents in the model, the focus is more on the heterogeneous household agents and their perceived risk during typhoon. The perceived risk of household agents which depends on three decision factors is the major determinant whether a household agent preemptively evacuates or not. Model simulation was performed using exhaustive exploration method on the set of input parameters. A single output variable determines the number of evacuation decisions.

The behavior of the agents was reviewed. It was seen that among the perceived risk decision factors, capacity-related factors contributes significantly in the decision of household agents to evacuate. The characteristics of the decision maker in a household and hazard-related factors also contributes but not as significant as capacity-related factors. Another deciding element observed is the threshold of evacuation decision. This threshold value is the assumed perceived limit of what a household decision maker can hold before finally deciding that the best action to do is to evacuate. These findings could provide insights to disaster managers and health emergency officers as to why households decides to preemptively evacuate while others opt to stay and wait for weather developments. It could also give initial understanding on how perceived risk and its decision factors directly affects how people behave during disaster such as typhoon.


\addtolength{\textheight}{-12cm}   



\section*{Acknowledgement}
The authors would like to thank the Municipal Disaster Risk Reduction and Management Office (MDRRMO) of LGU Bulan, Sorsogon for providing the necessary information used in this study and the Ateneo Social Computing Science Laboratory, Ateneo de Manila University for providing resources for the conduct of this study.

\bibliographystyle{unsrtnat}
\bibliography{references}  

\end{document}